\documentclass[english,twoside,a4paper,10pt]{article}

\usepackage[latin1]{inputenc}
\usepackage[T1]{fontenc}
\usepackage{amsmath}
\usepackage{amsfonts}
\usepackage{graphicx}
\usepackage{subfigure}
\usepackage{a4wide}
\usepackage{amssymb}
\usepackage{fancyhdr}
\usepackage{mathrsfs}
\usepackage[toc,page]{appendix}

\begin{document}

\title{\bf $f(\phi) R$-models for inflation}
\author{ 
Shynaray Myrzakul\footnote{Email: shynaray1981@gmail.com},\,\,\,
Ratbay Myrzakulov\footnote{Email: rmyrzakulov@gmail.com},\,\,\,
Lorenzo Sebastiani\footnote{E-mail address: l.sebastiani@science.unitn.it
}\\
\\
\begin{small}
Department of General \& Theoretical Physics and Eurasian Center for
\end{small}\\
\begin{small} 
Theoretical Physics, Eurasian National University, Astana 010008, Kazakhstan
\end{small}\\
}

\date{}

\maketitle

%%%%%%%%%%%%%%%%%%%%%%%%%%%%%%%%%%%%%%%%%%%%%%%%%%%%%%%%%%%%%%%%%%%%%%%%%%%%%%%%%%%%%%%%%%%%%%%%%%%%%%%%%%%%%%%%%%%%%%%%%%%%%%%%%%%

%%%%%%%%%%%%%%%%%%%%%
%  Abstract
%%%%%%%%%%%%%%%%%%%%%
\begin{abstract}
In this paper, we investigate models where a scalar field driving inflation is non minimally coupled with gravity and it is subjected to a scalar potential. We present several examples of coupling between the field and gravity, and we furnish realistic models for inflation in agreement with the last Planck results.
\end{abstract}
%%%%%%%%%%%%%%%%%%%%%

%----------------------------
%PACS
%----------------------------

%===========================================================================

%\tableofcontents
%%%%%%%%%%%%%%%%%%%%%%%%%%%
%%%  Sec. I
%%%%%%%%%%%%%%%%%%%%%%%%%%%
\section{Introduction}

The interest in the early-time expansion of our universe has grown considerably during the last years, following the accurate data coming from the cosmological observations~\cite{WMAP, Planck2}. The fact that the universe underwent an accelerated expansion after the Big Bang was first proposed by Guth~\cite{Guth} and Sato~\cite{Sato} in 1981 in order to explain the thermalization of the observable universe and in order to solve some problems related with the initial conditions of the Friedmann universe. Consequently, among the physicist community, it was understood that the inflationary paradigm also allows to produce the primordial perturbations at the origin of the galactic
structures, and the reproduction of such perturbations in agreement with observations is one of the most solid and fundamental test for every inflationary model (for reviews about inflation see Refs.~\cite{Linde, revinflazione} and references therein).

The number of models able to reproduce the early-time acceleration and the further
decelerated expansion of the Friedmann universe
is quite large. The classical picture is also known as ``chaotic inflation''  and is based on a scalar field, the inflaton, subjected to some potential~\cite{chaotic, buca1, buca2, buca3, buca4}. Typically, for large and negative values of the field, the potential is also very large 
and brings to a (quasi) de Sitter expansion where the curvature is near to the Planck scale. Therefore, the field slowly moves through a minimum of the potential where inflation ends and the reheating mechanism for the particle production takes place: alternatively, in the so called ``warm inflation scenario'', the radiation naturally arises at the end of inflation and the reaheating is not required~\cite{warm1, warm2, warm3, warm4, miowarm}. 
An other popular approach to inflation is furnished by the modified theories of gravity~\cite{reviewmod, Caprev, myrev}, where higher-derivative curvature corrections to the Einstein's theory may be at the origin of the early-time acceleration (see Refs.~\cite{Odinfrev, myrevinfl} for small review and Refs.~\cite{Staro, app1, app2, app3} for some applications).  

In this paper, we would like to consider a class of models where the field of inflaton with potential is minimally coupled with gravity during inflation (see Refs.~\cite{d1,d2} or Refs.~\cite{phi1,phi2,phi3} for recent works). 
We propose a brief investigation of such inflationary models. We note that the first models of chaotic inflation based on power-law potentials seem to be invalideted by the last cosmological data, and the study of scalar-field inflation minimally coupled with gravity with new mechanisms to produce the acceleration and the spectrum of primordial perturbations may bring to some valid alternative theory.
Different kinds of coupling between the field and gravity will be considered and discussed in the context of viable inflation, namely the spectral index and the tensor-to-scalar ratio of the models will be computed and compared with the Planck satellite surveys. 

The paper is organized as follows. In Section {\bf 2} we will present the formalism of the theory: a scalar field inducing inflation is subjected to a potential and it is minimally coupled with gravity. We will write the equations of motion and the conservation law of the field, and we will introduce the slow-roll parameters and the spectral index and the tensor-to-scalar ratio for such a kind of theory. Sections {\bf 3}--{\bf 4} are devoted to the investigation of different forms of coupling between the field and gravity, namely the exponential coupling and the power-law coupling. For every case, we will furnish viable models reproducing inflation according with the last Planck results. Final remarks are given in Section {\bf 5}.

%%% Unit %%%
We use units of $k_{\mathrm{B}} = c = \hbar = 1$ and denote the
gravitational constant, $G_N$, by $\kappa^2\equiv 8 \pi G_{N}$, such that
$G_{N}^{-1/2} =M_{\mathrm{Pl}}$, $M_{\mathrm{Pl}} =1.2 \times 10^{19}$ GeV being the Planck mass.
%%%%%%%%%%%%
%%%%%

\section{Formalism}

In this work, we consider a special class of $f(R,\phi)$-gravity, whose action reads
\begin{equation}
I=\int_{\mathcal M}d^4 x \sqrt{-g}\left[\frac{f(\phi) R}{2}-\frac{g^{\mu\nu}\partial_\mu\phi\partial_\nu\phi}{2}-V(\phi)
\right]\,,
\label{action}
\end{equation}
where $\mathcal M$ is the space-time manifold, $g$ is the determinant of the metric tensor $g_{\mu\nu}$ and $R$ is the Ricci scalar. The scalar field $\phi$ is subjected to the potential $V(\phi)$ and $f(\phi)$ represents a non-minimal coupling of the scalar field with gravity. When $f(\phi)=1/\kappa^2\equiv M_{Pl}^2/(8\pi)$, $M_{Pl}$ being the Planck mass, we recover the Einstein's framework.

In a flat Friedmann-Robertson-Walker (FRW) space-time,
\begin{equation}
ds^2=-dt^2+a(t)^2 {\bf dx}^2\,,\label{metric}
\end{equation}
with $a\equiv a(t)$ the scale factor,
the Equations of Motion (EOMs) of the theory are derived as
\begin{equation}
3 f(\phi) H^2=\frac{\dot\phi^2}{2}+V(\phi)-3 H\dot f(\phi)\,,\label{EOM1}
\end{equation}
\begin{equation}
-2 f(\phi)\dot H=\dot\phi^2+\ddot f(\phi)-H \dot f(\phi)\,,\label{EOM2}
\end{equation}
where $H=\dot a/a$ is the Hubble parameter and the dot denotes the time derivative. By combining this expressions, we also obtain the continuity equation of the scalar field, namely
\begin{equation}
\ddot\phi+3H\dot\phi-\frac{R}{2}\frac{d f(\phi)}{d\phi}+\frac{d V(\phi)}{d\phi}=0\,,\quad R=12H^2+6\dot H\,.\label{cons}
\end{equation}
During inflation the universe undergoes a phase of accelerated (quasi) de Sitter expansion, when the curvature (and therefore the Hubble parameter) is almost a constant and 
the magnitude of the
``slow-roll'' parameters~\cite{sr},
\begin{equation}
\epsilon_1=-\frac{\dot H}{H^2}\,,\quad\epsilon_2=\frac{\ddot\phi}{H\dot\phi}\,,\quad
\epsilon_3=\frac{\dot f(\phi)}{2 H f(\phi)}\,,\quad
\epsilon_4=\frac{\dot E}{2 H E}\,,\label{srpar}
\end{equation}
with
\begin{equation}
E=f(\phi)+\frac{3\dot f(\phi)^2}{2\dot\phi^2}\,,
\end{equation}
is extremelly small. Thus, in the so called ``slow-roll approximation'', Eq.~(\ref{EOM1}) with Eq.~(\ref{cons}) read
\begin{equation}
3f(\phi) H^2\simeq V(\phi)\,,\quad
3H\dot\phi-6H^2\frac{d f(\phi)}{d\phi}+\frac{d V(\phi)}{d\phi}
\simeq 0\,.\label{EOMsr}
\end{equation}
In order to measure the spectrum of the perturbations during inflation, we need the spectral index $n_s$ and the tensor-to-scalar ratio $r$~\cite{sr, corea},
\begin{equation}
n_s=1-4\epsilon_1-2\epsilon_2+2\epsilon_3-2\epsilon_4\,,\quad
r=16(\epsilon_1+\epsilon_3)\,,\label{ind}
\end{equation}
where $|\epsilon_{1,2,3,4}|\ll 1$ must be evaluated in the slow-roll regime\footnote{Note that in scalar tensor theory with $f(\phi)=1$, $\epsilon_2=\epsilon_1-\eta$, where $\eta=\epsilon_1-\ddot H/(2H\dot H)$, such that $n_s=1-6\epsilon_1+2\eta$ and $r=16\epsilon_1$.}. 

The following relations hold true~\cite{phi2},
\begin{equation}
\epsilon_4=
\frac{\left[\frac{-4\dot\phi^2}{H\dot f(\phi)}\epsilon_3
+6\epsilon_1+6\epsilon_3(1-\epsilon_2)
\right]}{2\left[\frac{\dot\phi^2}{H\dot f(\phi)}+3\epsilon_3\right]}\,,\nonumber
\end{equation}
\begin{equation}
\epsilon_1=-\epsilon_3+\frac{\dot\phi^2}{3H\dot f(\phi)}\left(\epsilon_4+2\epsilon_3\right)+\epsilon_3(\epsilon_2+\epsilon_4)\simeq
-\epsilon_3+\frac{\dot\phi^2}{3H\dot f(\phi)}\left(\epsilon_4+2\epsilon_3\right)\,,
\end{equation}
such that the spectral index and the tensor-to-scalar ratio in (\ref{ind}) can be rewritten as
\begin{eqnarray}
n_s&\simeq& 1-2\epsilon_1\left(\frac{3H\dot f(\phi)}{\dot\phi^2}+2\right)-2\epsilon_2-6\epsilon_3\left(\frac{H\dot f(\phi)}{\dot\phi^2}-1\right)\nonumber\\
&=& 1+
\frac{2\dot H}{H^2}\left(\frac{3H\dot f(\phi)}{\dot\phi^2}+2\right)
-\frac{\ddot\phi}{H\dot\phi}-\frac{3\dot f(\phi)}{H f(\phi)}\left(\frac{H\dot f(\phi)}{\dot\phi^2}-1\right)\,,\label{index2}
\\
r&=&16(\epsilon_1+\epsilon_3)=-\frac{16\dot H}{H^2}+\frac{8\dot F(R,\phi)}{H F(R,\phi)}\,.\label{ratio2}
\end{eqnarray}
The last cosmological Planck satellite data~\cite{Planck2} constrain these two quantities as
$n_{\mathrm{s}} = 0.968 \pm 0.006\, (68\%\,\mathrm{CL})$ and 
$r < 0.11\, (95\%\,\mathrm{CL})$.

Finally, the total amount of inflation is measured by the $e$-folds number
\begin{equation}
\mathcal N\equiv\ln \left(\frac{a_\mathrm{f}(t_\text{f})}{a_\mathrm{i} (t_\text{i})}\right)=\int^{t_\text{f}}_{t_\text{i}} H(t)
dt\,,\label{Nfolds}
\end{equation}
$a_\text{i}(t_{i})$ and $a_{\text{f}}(t_{\text{f}})$ being the scale factor at the beginning and at the end of inflation, respectively, and $t_\text{i,f}$ the related times. To have the thermalization of our observable universe, it must be  at least $\mathcal N\simeq 60$. 

In the next sections, we will analyze different way to obtain viable inflation from this kind of theories.

%%%%%%%%%%%%%%%%%%%%%%%%%%%%%%%%%%%%%%%%%%%%%%%%%%%%%%%%%%%%%%%%%%%%%%%%%%%%%%%%%%%%%%%%%%%%%%%%%%%%%%%%%%%%%%%%%%%%%%%%%%%%%%%%%%%

\section{Exponential coupling between the field and gravity}

At first, we explore the following form of non-minimal coupling between the field and gravity,
\begin{equation}
f(\phi)=\frac{\text{e}^{n\kappa\phi}}{\kappa^2}\,,\quad 0<n\,,
\end{equation} 
where $n$ is a positive parameter and the Plank Mass encoded in $\kappa^2\equiv 8\pi/M_{Pl}^2$ has been introduced for dimensional reasons. 
Inflation is realized  for large and negative values of the field, and in order to 
get the de Sitter expansion
we need the following suitable form for the field potential,
\begin{equation}
V(\phi)=c_0\left(1-\text{e}^{n\kappa\phi}\right)\,,
\end{equation}
where $c_0$ is a constant with dimension $[c_0]=[1/\kappa^4]$. In this way, the potential tends to a constant during inflation, namely $V(\phi\rightarrow-\infty)\simeq c_0$, but, when the field is close to zero at the end of inflation, it falls in the minimum of the potential located at $V(\phi=0)=0$ and, since $f(\phi=0)=1$, we may recover the Friedmann universe, eventually after the reheating processes.

When the field is negative and very large\footnote{Note that the field can exceed the Planck mass when $0<n<1$.},
\begin{equation}
\frac{1}{n\kappa}\ll|\phi|\,,\label{cond0}
\end{equation}
the EOMs (\ref{EOM1})--(\ref{EOM2}) are simplified as in (\ref{EOMsr}) and one has
\begin{equation}
H^2\simeq\frac{\kappa^2 c_0}{3}\text{e}^{-n\kappa\phi}\,,\quad \dot\phi\simeq
\frac{2c_0 n\kappa}{3H}\,.
\end{equation}
We see that when $\phi\rightarrow-\infty$ the Hubble parameter is very large and $\kappa^2 c_0\ll H^2$, such that the field slowly
increases during inflation. It is easy to find
\begin{equation}
\dot f(\phi)=(n\kappa\dot\phi)f(\phi)\,,\quad
\ddot f(\phi)=(n\kappa\ddot\phi)f(\phi)+(n\kappa\dot\phi)^2 f(\phi)\,,\quad
\ddot\phi=-\frac{2 c_0 n\kappa\dot H}{3H^2}\,.
\end{equation}
Since
\begin{equation}
\dot H=-\frac{(c_0 n\kappa)^2}{3c_0}\,,
\end{equation}
which is consistent with (\ref{EOM2}) in slow-roll approximation, from (\ref{srpar}) we get
\begin{equation}
\epsilon_1=n^2\text{e}^{n\kappa\phi_0}\,,\quad
\epsilon_2=
\epsilon_3=\epsilon_1\,,
\end{equation}
where $\phi_0$ is the value of the field at the beginning of inflation: when $\phi_0\rightarrow-\infty$, the slow-roll parameters are small and the slow-roll approximation holds true. On the other hand, when the slow-roll parameters are of the order of the unit (in particular, when $\epsilon_1\simeq 1$ such that $|\dot H|\simeq H^2$), the slow-roll approximation is not still valid and the universe exits from the de Sitter accelerated expansion.
Now, by taking into account that $H\dot f(\phi)/\dot\phi^2=1/2$, we proceed to calculate the spectral index and the tensor-to-scalar ratio from (\ref{index2})--(\ref{ratio2}),
\begin{equation}
\left(1-n_s\right)
=
6\epsilon_1
\equiv 6n^2\text{e}^{n\kappa\phi_0}
\,,\quad
r=32\epsilon_1\equiv 32 n^2\text{e}^{n\kappa\phi_0}\,.\label{sr1}
\end{equation}
Moreover, the $e$-folds in (\ref{Nfolds}) reads
\begin{equation}
\mathcal N=\int^{t_\text{f}}_{t_{\text{i}}}H dt=\int^{\phi_\text{f}}_{\phi_0}\frac{H}{\dot\phi} d\phi=
\int^{\phi_\text{f}}_{\phi_0}\frac{\kappa\text{e}^{-n\kappa\phi}}{2n} d\phi
\simeq
\frac{\text{e}^{-n\kappa\phi_0}}{2n^2}\,, 
\end{equation}
where we have taken into account that the argument of the integral finally depends on $\phi$ only and $|\phi_\text{f}|\ll |\phi_0|$, $\phi_\text{f}$ being the value of the field at the end of inflation. Thus, by plugging this expression into (\ref{sr1}), we get
\begin{equation}
\left(1-n_s\right)= \frac{3}{\mathcal N}\,,\quad
r=\frac{16}{\mathcal N}\,.\label{indexes1}
\end{equation}
This result does not depend on the parameter $n$. For $\mathcal N\simeq 80$, one may recover a spectral index in agreement with the last Planck results, but the tensor-to-scalar ratio of the model is too large.\\
\\
An other exponential coupling between the field and gravity that we would like to consider in our investigation assumes the following form,
\begin{equation}
f(\phi)=\frac{1+\text{e}^{n\kappa\phi}}{2\kappa^2}\,,\quad 0<n\,.
\end{equation}
In this case, $f(\phi\rightarrow-\infty)\simeq 1/(2\kappa^2)$ during inflation, but when $\phi$ is close to zero we still recover the result of General Relativity with $f(\phi=0)=1/\kappa^2$. For our aim, we will make use of a constant potential, 
\begin{equation}
V(\phi)=c_0\,.
\end{equation}
It is understood that it is possible to add any suitable dynamical part to the potential, under the requirement that it changes much slower respect to $f(\phi)$: for example, with a potential $V(\phi)=c_0(1-\exp\left[2n\kappa\phi\right])$, we have $V(\phi=0)=0$ at the end of inflation, but the dynamics of the field in the slow-roll approximation is determined by $f(\phi)$ only  and can be analyzed as in the following. Analogue models can be derived by considering more general cases $f(\phi)=\left(a+(2-a)\exp\left[2n\kappa\phi\right]\right)/(2\kappa^2)$, with $a$ real number.

In the limit (\ref{cond0}), we obtain the (quasi) de Sitter solution, 
\begin{equation}
H^2\simeq\frac{2c_0\kappa^2}{3}\,,\quad\dot\phi\simeq\frac{H n\text{e}^{n\kappa\phi}}{\kappa}\,,
\end{equation}
such that the field moves slowly during inflation. With a similar procedure of the preceeding example, the slow-roll parameters  can be derived as ($\phi_0$ is the value of the field at the beginning of inflation),
\begin{equation}
\epsilon_1=\frac{n^2\text{e}^{2n\kappa\phi_0}}{2}
\,,\quad
\epsilon_2=n^2\text{e}^{n\kappa\phi_0}+\epsilon_1\,,\quad
\epsilon_3=
\frac{n^2\text{e}^{2n\kappa\phi_0}}{2+2\text{e}^{n\kappa\phi_0}}\simeq\epsilon_1\,.
\end{equation}
Since $H\dot f(\phi)/\dot\phi^2=1/2$, one has
(in the slow-roll limit $\phi_0\rightarrow-\infty$),
\begin{equation}
(1-n_s)\simeq 2n^2\text{e}^{n\kappa\phi_0}
\,,
\quad r\simeq 16n^2\text{e}^{2n\kappa\phi_0}
\,.
\end{equation}
The $e$-folds is given by
\begin{equation}
\mathcal N=
\int^{\phi_\text{f}}_{\phi_0}\frac{\kappa\text{e}^{-n\kappa\phi}}{n} d\phi
\simeq
\frac{\text{e}^{-n\kappa\phi_0}}{n^2} \,,
\end{equation}
and finally we can write
\begin{equation}
(1-n_s)\simeq\frac{2}{\mathcal N}\,,\quad r\simeq\frac{16}{n^2 N^2}\,.
\end{equation}
Thus, in order to recover the spectral index inferred from the Planck data, it is enough an amount of inflation $\mathcal N\simeq 60$, while the tensor-to-scalar ratio of the model depends on the parameter $n$ and in general results to be extremelly small according with observations.
In particular, we note that, if we choose $n=2/\sqrt{3}$, the model leads to the same indexes of the so called ``Starobinsky model'', where a correction quadratic in the Ricci scalar of the action of General Relativity supports inflation at high curvature~\cite{Staro}.

\section{Power-law coupling between the field and gravity}

As a second class of models, we will investigate the following form of coupling between the filed and gravity,
\begin{equation}
f(\phi)=\frac{1}{\kappa^2}\left(\frac{\phi_\text{f}}{\phi}\right)^n\,,\quad 0<n\,,
\end{equation}
where $\phi_\text{f}$ is the value of the field at the end of inflation, when one can recover the Hilbert-Einstein action of General Relativity with $f(\phi)\simeq1/\kappa^2$. We use the potential 
\begin{equation}
V(\phi)=c_0\left(1-\left(\frac{\phi_\text{f}}{\phi}\right)^n\right)\,,
\end{equation}
with $V(\phi=\phi_{\text{f}})= 0$ at the end of inflation. Thanks to the coupling function and the potential above, in the limit $\phi\rightarrow-\infty$ the space-time is described by the (quasi) de Sitter solution
\begin{equation}
H^2\simeq\frac{c_0 \kappa^2}{3}\left(\frac{\phi}{\phi_\text{f}}\right)^n\,,\quad
\dot\phi\simeq-\frac{2c_0 n }{3 H \phi}\,.
\end{equation}
Thus, the field slowly increases during inflation. At the same time, the Hubble parameter decreases as
\begin{equation} 
\dot H\simeq -\frac{c n^2}{3\phi^2}\,,
\end{equation}
and, when its magnitude is on the same order of $H^2$, the universe exits from the de Sitter phase. 
In the slow-roll approximation with $\phi_0\rightarrow-\infty$, the slow-roll parameters in (\ref{srpar}) result to be
\begin{equation}
\epsilon_1=\frac{n^2}{\kappa^2\phi_0^2}\left(\frac{\phi_\text{f}}{\phi_0}\right)^n\,,\quad
\epsilon_2=\epsilon_1\left(1+\frac{2}{n}\right)\,,\quad
\epsilon_3=\epsilon_1\,,
\end{equation}
with $\phi_0$ the boundary value of the field at the beginning of inflation. Now, by using $H\dot f(\phi)/\dot \phi^2=1/2$ in (\ref{index2})--(\ref{ratio2}), we get
\begin{equation}
(1-n_s)\simeq\frac{2n(2+3n)}{\kappa^2\phi^2}\left(\frac{\phi_\text{f}}{\phi}\right)^n\,,
\quad
r_s\simeq\frac{32 n^2}{\kappa^2\phi^2}\left(\frac{\phi_\text{f}}{\phi}\right)^{n}\,.
\end{equation}
The $e$-folds (\ref{Nfolds}) of the model can be easily evaluated as
\begin{equation}
\mathcal N=
-\int^{\phi_\text{f}}_{\phi_0} \frac{\kappa^2\phi}{2n}\left(\frac{\phi}{\phi_\text{f}}\right)^n d\phi
\simeq
\frac{\kappa^2\phi_0^2}{2n(n+2)}\left(\frac{\phi_0}{\phi_\text{f}}\right)^n\,,
\end{equation}
and implies
\begin{equation}
(1-n_s)\simeq\frac{(2+3n)}{(2+n)\mathcal N}\,,\quad r\simeq\frac{16 n}{(2+n)\mathcal N}\,.
\end{equation}
If we set $n=2$ we obtain,
\begin{equation}
(1-n_s)\simeq\frac{2}{\mathcal N}\,,\quad
r_s\simeq\frac{8}{\mathcal N}\,,\quad n=2\,,
\end{equation} 
namely the indexes of the power-law chaotic inflation with quadratic potential. This kind of inflation does not realize the tensor-to-scalar ratio inferred from the Planck observations. To do it, we need $2/3<n<2$.\\
\\
As a final example, we will consider the following coupling between the field and gravity,
\begin{equation}
f(\phi)=\frac{\left(1+\left(\frac{\phi_\text{f}}{\phi}\right)^n\right)}{2\kappa^2}\,,\quad 0<n\,.
\end{equation}
Also in this case, when the field reaches the value $\phi_\text{f}$ at the end of inflation, one recovers the gravitational lagrangian of General Relativity with $f(\phi=\phi_\text{f})=1/\kappa^2$.

Let us take the following potential,
\begin{equation}
V(\phi)=c_0\left(\frac{\phi}{\phi_0}\right)^m\,,\quad 0<m\,,
\end{equation}
with $m$ positive parameter. When the field is very large and assumes the value $\phi_0$ at the beginning of inflation, the potential tends to a constant, namely $V(\phi=\phi_0)=c_0$, but when the field goes to zero at the end of inflation\footnote{At the end of inflation $\phi=\phi_\text{e}$ with $\phi_\text{e}\rightarrow 0^-$.} the potential vanishes, namely $V(\phi=0)=0$.
The (quasi) de Sitter solution and the time derivative of the field during inflation are derived from Eq.~(\ref{EOMsr}) in the limit $\phi\rightarrow\phi_0$, with $\phi_0\rightarrow-\infty$, namely
\begin{equation}
H^2\simeq\frac{2c_0\kappa^2}{3}\left(\frac{\phi}{\phi_0}\right)^m\,,\quad\dot\phi\simeq-\frac{c_0 m}{3H\phi}
\left(\frac{\phi}{\phi_0}\right)^m\,.
\end{equation}
Thus, the field slowly increases during inflation and the Hubble parameter evolves as
\begin{equation}
\dot H\simeq-\frac{c_0 m^2}{6\phi^2}\,.
\end{equation}
For this model the slow-roll parameters read, in the slow roll limit $\phi\rightarrow\phi_0$ and $|\phi_\text{f}|\ll |\phi_0|$,
\begin{equation}
\epsilon_1\simeq\frac{m^2}{4\kappa^2\phi_0^2}\,,\quad
\epsilon_2\simeq\frac{m(2-m)}{4\kappa^2\phi_0^2}\,,\quad
\epsilon_3\simeq\frac{m n}{4\kappa^2\phi_0^2}\left(\frac{\phi_\text{f}}{\phi_0}\right)^n\,.
\end{equation}
Differently from the other examples analyzed in this paper, here one has 
$H\dot f(\phi)/\dot \phi^2=(n/m)(\phi_\text{f}/\phi_0)^n\ll 1$, such that we can rewrite (\ref{index2})--(\ref{ratio2})
as
\begin{eqnarray}
n_s\simeq 1-4\epsilon_1-2\epsilon_2+6\epsilon_3
\,,\quad
r=16\epsilon_1\,,
\end{eqnarray}
which lead to
\begin{equation}
(1-n_s)\simeq\frac{m(m+2)}{2\kappa^2\phi_0^2}\,,\quad
r\simeq
\frac{m^2}{\kappa^4\phi_0^2}\,.
\end{equation}
The $e$-folds number (\ref{Nfolds}) reads
\begin{equation}
\mathcal N=
-\int^{\phi_\text{f}}_{\phi_0} \frac{2\kappa^2\phi}{m} d\phi
\simeq
\frac{\kappa^2\phi_0^2}{m}\,.
\end{equation}
As a result,
\begin{equation}
(1-n_s)\simeq\frac{(2+m)}{2\mathcal N}\,,\quad
r\simeq\frac{m}{\mathcal N}\,,
\end{equation}
and we see that we can satisfy the Planck data when $\mathcal N\simeq 60$ for $m=2$ or, if we consider $\mathcal N\simeq 80$, for $m=4$. We may conclude that this model leads to realistic inflationary scenario only if $2<m<4$ with $e$-folds $60<\mathcal N<80$.

\section{Conclusions}

In this paper, we considered models for inflation where a scalar field (also dubbed ``inflaton'') is minimally coupled with gravity and it is subjected to a scalar potential. Inflation from scalar field is quite popular and is well-known in literature. In the Einstein's framework (in the absence of the coupling between the field and gravity) scalar field with exponential potential is one of the most realistic choice to reproduce (chaotic) inflation: we remember that also inflation supported by quadratic corrections of the Ricci scalar to the lagrangian of General Relativity, after a conformal transformation, can be reduced to this kind of theory. On the other side, other forms of potential seem to do not reproduce the last Planck results: for example, potentials quadratic in the scalar field lead to a tensor-to-scalar ratio slightly bigger than the observed one, while with larger power-law potentials the spectral index results to be too close to one. In this respect, the presence of a coupling between the field and gravity may bring to 
interesting results for the early-time acceleration. 

The models analyzed in this paper are quite simple and lead to realistic scenarios for early-time universe. In general, the coupling between the field and gravity vanishes at the end of inflation, when the Friedmann universe has to be recovered. We considered two classes of coupling, namely exponential coupling and power-law coupling. The spectral index and the tensor-to-scalar ratio have been explicitly calculated as functions of the $e$-folds number for every model. This results can be easily compared with the last Planck satellite data and we found the way to reproduce a realistic scenario in the two considered classes of models. In particular we note that also in the presence of power-law potentials, one can get viable inflation thanks to the coupling between the field and gravity.

Due to the fact that inflation occurs at high energy, it is expected that quantum corrections appear in the the theory. In this respect, more general approaches must be considered (on the other hand, the freedom degree introduced by a scalar field may have a correspondance in terms of curvature invariants). In the specific, we could make use of functions of other curvature invariants (in particular, the contraction of the Weyl tensor and the Gauss-Bonnet) to describe the gravitational Lagrangian. As an interesting example, in Ref.~\cite{RGinfl} the authors explored the possibility to recover the early-time acceleration in the framework of $f(R,G)$-gravity, $G$ being the Gauss Bonnet four dimensional topological invariant, and used the different scales of the two curvature invariants to reproduce a double inflation scenario: the power models there considered can describe viable inflation.

\end{document}